\documentclass[aps,twocolumn]{revtex4}
\usepackage[dvips]{graphics,graphicx}
\usepackage{amsmath}

\usepackage{amssymb}
\usepackage{epstopdf}
\usepackage[usenames, dvipsnames]{color}
\usepackage{graphicx}
\usepackage{graphics}
\usepackage{amsfonts}
\usepackage{eufrak}
\usepackage{mathrsfs}
\usepackage{bm}
\newcommand{\scs}{\scriptscriptstyle}

\begin{document}

\title{Resonant transmission of fermionic carriers: comparison between solid-state physics and quantum optics approaches}
\author{ Andrey~R.~Kolovsky$^{1,2}$ and Dmitrii~N.~Maksimov$^{1,3}$}
\affiliation{$^1$Kirensky Institute of Physics, Federal Research Center KSC SB
RAS, 660036, Krasnoyarsk, Russia}
\affiliation{$^2$School of Engineering Physics and Radio Electronics,
Siberian Federal University, 660041, Krasnoyarsk, Russia}
\affiliation{$^3$IRS SQC, Siberian Federal University, 660041, Krasnoyarsk, Russia}

\date{\today}
\begin{abstract}
We revisit the phenomenon of the resonant transmission of fermionic carriers  through a quantum device  connected to two contacts with  different chemical potentials. We show that, besides the traditional in solid-state physics Landauer-B\"uttiker approach, this phenomenon can be also described  by the non-Markovian master equation for the reduced density matrix of the fermions in the quantum device. We identify validity regions of both approaches in the system parameter space and argue that for large relaxation rates the accuracy of the latter approach greatly exceeds the accuracy of the former.
\end{abstract}
\maketitle

\section{Introduction}
\label{sec1}

The problem of electron transport in a metallic wire connecting two contacts is older than quantum mechanics \cite{Ohm27}. In the past centuries this problem was readdressed several times reflecting the progress in experimental physics, where the main milestones are the appearance of clean semiconductors and lithography technology and the emergence of physics of ultra-cold atoms. The former technology  substituted the wire with an engineered device -- a quantum dot \cite{Kast93,Asho96}, while with cold atoms in optical lattice one can mimic the behavior of crystalline electrons  in the pure form, i.e., without complications caused by the presence of the long-range Coulomb interaction and electron-phonon interaction \cite{ Seam03,Bloc12}. These two systems -- quantum dots and cold atoms  -- allow experimentalists to study the coherent transport of carriers (electrons and neutral atoms, respectively) where deviations from the classical Ohm law become especially pronounced.

As concerns the theory, presently we have a vast variety of  methods which, however, can be sorted into two large groups. The methods belonging to the first group, which we shall refer to as the solid-state physics approach, are traced back to Landauer's conjecture \cite{Land92} that the wire conductance is related to the transmission probability  and, as a rule, they extensively use the Green function formalism. The famous result  of the solid-state physics approach is the theoretical description of the phenomenon of resonant transmission in quantum dots \cite{Datt95,Datt05}. The methods belonging to the second group, which we shall refer to as the quantum optics approach, operate with very different notions like the quantum master equation for open (generally, many-body) quantum systems and the Born and Markov approximations \cite{Davi76,Breu07,117,Vyas20,Land21}.  One can also assign to this group the stochastic methods which explore the correspondence between the master equations and the stochastic Schr\"odinger equations \cite{Dios98,Zhao12,Chen13}. Remarkably, in spite of the completely different technique the quantum optics approach is also capable to capture the phenomenon of the resonant transmission \cite{preprint}. The question arises of how the results of the above two approaches are related to each other and which of them is more accurate.  In the present work we answer this question by studying the simple model for quantum transport of fermionic carriers introduced in Ref.~\cite{120}. This model can be viewed  either as a generalisation of the open Hubbard models \cite{Pros14,Ivan13,112} onto the case of arbitrary reservoir temperature or as a generalisation of the Landauer approach for the electron transport \cite{Datt95,Datt05}  where the relaxation processes in the contacts are explicitly taken into account.  Thus, the model can be equally analysed by using both solid-state  physics and quantum optics approaches.

The structure of the paper is as follows. In Sec.~\ref{sec2} we recall the ingredients of the model and preliminary discuss the system dynamical regimes depending on the control parameter.  Analytical results are collected in Sec.~\ref{sec3}. This section consists of three subsections where we employ three different methods to study the system, namely, the Markovian master equation, non-Markovian master equation, and  the Landauer-like approach. In Sec.~\ref{sec4} we analyse the coherent properties of the carriers that help us to quantify the degree of validity of the used approaches. The main results of the work are summarised in the concluding Sec.~\ref{sec5}.

\section{The model}
\label{sec2}

We consider the set-up consisting of a linear tight-binding chain of the length $L$ coupled at both ends with two tight-binding rings of $M$ sites each, see Fig.~1 in Ref.~\cite{120}. Throughout the text the rings are termed the contacts since they served as the particle reservoirs. Non-interacting spinless fermions can hop between the sites of the chain and the sites of the rings with the rates $J_{\rm s}$  and $J_{\rm r}$, where $J_{\rm s}\sim J_{\rm r}$, while the hopping between the chain and the contacts is quantified by the coupling constant $\epsilon\ll J_{\rm s}, J_{\rm r}$. The dynamics is governed by the master equation for the total density matrix
\begin{equation}\label{Master_full}
\frac{\partial \widehat{{\cal R}}}{\partial t}=-i[\widehat{{\cal H}},  \widehat{{\cal R}}]+
\gamma\sum_{\ell=1,L}\left(\widehat{{\cal L}}^{\scs (g)}_{\ell}+\widehat{{\cal L}}^{\scs (d)}_{\ell}\right).
\end{equation}
In Eq. (\ref{Master_full})  the Hamiltonian has the form
\begin{align}\label{Ham_tot}
\widehat{{\cal H}}=\widehat{{\cal H}}_{\rm s}+\sum_{\ell=1,L}\left(
\widehat{{\cal H}}_{{\rm r},\ell}+\widehat{{\cal H}}_{{\rm c},\ell}\right),
\end{align}
where
\begin{align}\label{Ham_sys}
\widehat{{\cal H}}_{\rm s}=
-\frac{J_{\rm s}}{2}\sum_{\ell=1}^{L-1}\hat{c}_{\ell+1}^{\dagger}\hat{c}_{\ell} +{\rm h.c.}
\end{align}
is the chain Hamiltonian with $\hat{c}_{\ell}^{\dagger},\hat{c}_{\ell}$ being Fermionic creation and annihilation operators at the $\ell$th site. The contact Hamiltonians,  $\widehat{{\cal H}}_{{\rm r},\ell}$,  and the coupling  Hamiltonians, $\widehat{{\cal H}}_{{\rm c},\ell}$, are indexed with subscript $\ell=1,L$ specifying the connection site. The contact Hamiltonians are written in terms of Fermionic operators acting in the Fock space of the Bloch eigenstates of the ring,
\begin{equation}\label{Ham_res}
\widehat{{\cal H}}_{\rm r}=
-J_{\rm r}\sum_{k=1}^{M}\cos\left(\frac{2\pi k}{M}\right) \hat{b}_{k}^{\dagger}\hat{b}_{k} 
\end{equation}
(here we dropped subscript $\ell$  assuming that the ring are identical). The coupling Hamiltonian is given by
\begin{align}\label{Ham_cou}
\widehat{{\cal H}}_{\rm c, \ell}=
-\frac{\epsilon}{2\sqrt{M}}\sum_{k=1}^M\hat{c}_{\ell}^{\dagger}\hat{b}_{k} +{\rm h.c.}.
\end{align}

\begin{figure}[t]
\includegraphics[width=8.5cm,clip]{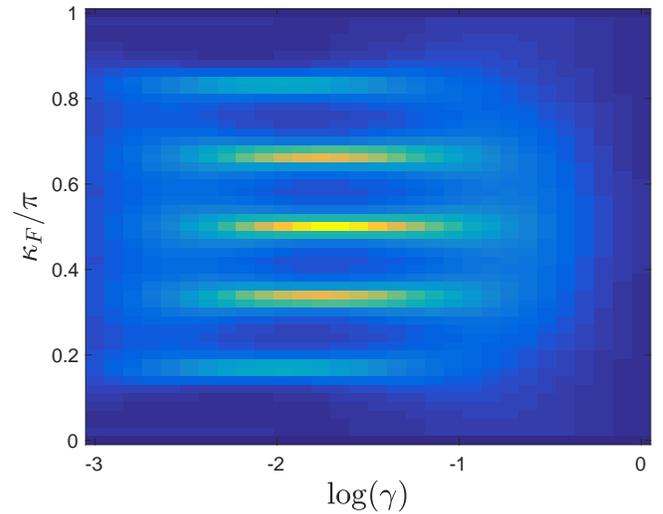}
\caption{The current across the tight-binding chain of the length $L=5$ as the function of $\kappa_F$ and the relaxation constant $\gamma$. The hopping matrix elements are $J_{\rm s}=J_{\rm r}=1$, the coupling constant $\epsilon=0.4$, the temperature $1/\beta=0$, the size of the rings $M=100$, and the difference in the contact chemical potentials $\Delta \mu=J_r\sin(\kappa_F)(2\pi/M)$.}
\label{fig1}
\end{figure} 

To prescribe thermodynamic quantities to each contact we introduced the particle drain,
\begin{equation}\label{drain}
\widehat{{\cal L}}^{\scs(d)}_{\ell}=\sum_{k=1}^M\frac{\bar{n}_{k,\ell}-1}{2}
\left(\hat{b}_{k}^{\dagger}\hat{b}_{k}\widehat{\cal R }-2\hat{b}_{k}\widehat{\cal R }\hat{b}_{k}^{\dagger}
+\widehat{\cal R }\hat{b}_{k}^{\dagger}\hat{b}_{k} \right) \;,
\end{equation}
and the particle gain,
\begin{equation}\label{gain}
\widehat{{\cal L}}^{\scs{(g)}}_{\ell}=-\sum_{k=1}^M\frac{\bar{n}_{k,\ell}}{2}
\left(\hat{b}_{k}\hat{b}_{k}^{\dagger}\widehat{\cal R }-2\hat{b}_{k}^{\dagger}\widehat{\cal R }\hat{b}_{k}
+\widehat{\cal R }\hat{b}_{k}\hat{b}_{k}^{\dagger} \right) \;,
\end{equation}
Lindblad operators, where
\begin{equation}
\label{Fermi}
\bar{n}_{k,\ell}= \frac{1}{e^{-\beta_{\ell}[J_{\rm r}\cos(2\pi k/M)+\mu_{\ell}]}+1} \;.
\end{equation}
This ensures that the Bloch states of  isolated contacts are populated according to the Fermi-Dirac distribution with given chemical potential $\mu_{\ell}$ and inverse temperature $\beta_{\ell}$. Finally, the constant $\gamma$ in Eq.~(\ref{Master_full}) is the  relaxation rate which determines how fast the isolated contacts relax to their thermodynamic equilibrium. It was shown in Ref.~\cite{120} that the considered model validates the main assumption of the Landauer approach about the reflectionless contacts, thus, allowing for the straightforward application of this theory.
\begin{figure}
\includegraphics[width=8.5cm,clip]{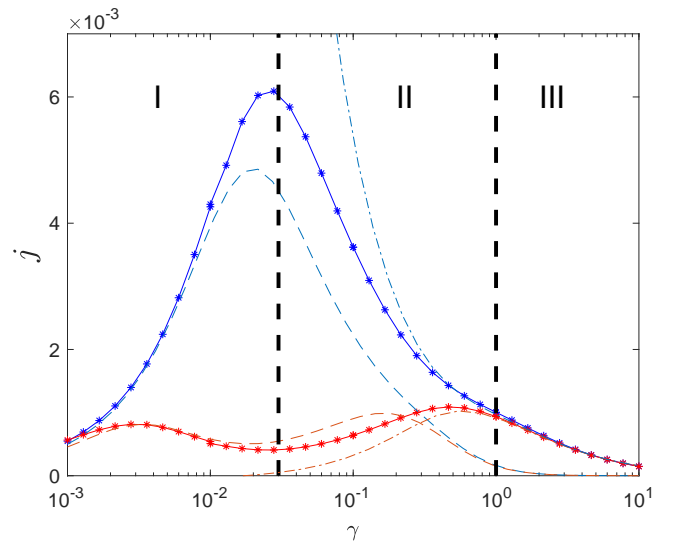}
\caption{The current as the function the relaxation constant $\gamma$  for $\kappa_F=\pi/2$ ($\mu=0$), blue asterisks, and $\kappa_F\approx0.58\pi$ ($\mu=0.25$), red circles. Vertical dashed lines demark different dynamical regimes of the system with respect to the Born and Markov approximations. The additional dash and dash-dotted lines are results of the algebraic approaches, see text.}
\label{fig2}
\end{figure}

Unlike Ref.~\cite{120} in the present work we shall consider short chain where one can observe the phenomenon of resonant transmission. Also we focus on the linear response regime where the current across the chain is proportional to the contact chemical potential difference.  As an example, Fig.~\ref{fig1}  and Fig.~\ref{fig2} show the results of numerical analysis of the model for $\beta=\infty$, $L=5$ and $M=100$, which is large enough to speak about quasi-continuous spectrum of the contacts. In these numerical simulations we evolve the system in time till the density matrix ${\cal R}(t)$ reaches its stationary value from which we determine the stationary current $\langle j\rangle$. In Fig.~\ref{fig1} we depict the 
stationary current
as a color map where the phenomenon of the resonant transmission is seen as local peaks at the values of the chemical potential $\mu=-J_r\cos(\kappa_F)$ coinciding with eigenenergies of the isolated tight-binding chain. Notice that the peaks are present only in certain interval of $\gamma$.  Fig.~\ref{fig2} shows the current as the function of $\gamma$ for $\mu=0$ (the central peak in Fig.~\ref{fig1}) and $\mu=0.25$ (the nearby valley). This figure is aimed to illustrate the Esaki-Tsu-like dependence \cite{Esak70,Mino04,Ott04,77} of the current on the relaxation rate $\gamma$, with the universal asymptotic $j \sim\gamma$ for $\gamma\rightarrow0$ and $j \sim1/\gamma$ for $\gamma\rightarrow\infty$. In Fig.~\ref{fig2} we also demark by the vertical dashed lines the different dynamical regimes of the system which we briefly discuss in the  next paragraph.

In parameter region III the large value of the relaxation constant $\gamma$ validates the Born and Markov approximations that allow us to derive  the Markovian master equation for the reduced density matrix of the carriers in the chain, see Sec.~\ref{sec3a}. In region II the Markov approximation fails, however, the Born approximation is still valid. In the other words, the ergodic properties of the contacts, when they are viewed as particle reservoirs, are not affected by the presence of the chain. The direct consequence of this fact is that the reduced density matrices of the contacts remain close to their  thermodynamic equilibrium \cite{remark1}. The failure  of the Markov approximation leads  to the non-Markovian   (integrodifferential) master equation for the reduced density matrix of the carriers in the chain, see Sec.~\ref{sec3b}. Finally, if we move to region I we break both the Markov and Born approximations.  To analyse the system in this region we employ the solid-state physics approach. It is shown in Sec.~\ref{sec3c} that this approach gives quantitatively correct results in region I and qualitatively correct results in region II. However, it fails to reproduce the case of large $\gamma$ where it gives the wrong asymptotic $ \langle j\rangle\sim 1/\gamma^2$.

\section{Quantum transport across the tight-binding chain}
\label{sec3}

Eq.~(\ref{Master_full}) contains only pairwise combinations of the creation and annihilation operators. This allows us to rewrite it in terms of the single particle density matrix (SPDM) $\hat{\rho}$. Let us assume for a moment that only one (the left) contact is attached to the chain. Then the total SPDM takes the following block form,
\begin{equation}\label{SPDM_block}
\hat{\rho}=
\left(
\begin{array}{cc}
\hat{\rho}_{\rm r} & \hat{\rho}_{\rm c} \\
\hat{\rho}_{\rm c}^{\dagger} & \hat{\rho}_{\rm s}
\end{array}
\right),
\end{equation}
where $\hat{\rho}_{\rm r}$ is the SPDM of the contact with the elements  $\rho_{k,k'}={\rm Tr}(\hat{b}_k^\dagger\hat{b}_{k'} \hat{\cal{R}})$, 
 $\hat{\rho}_{\rm s}$ is the SPDM of the chain with the elements  $\rho_{\ell,\ell'}={\rm Tr}(\hat{c}_\ell^\dagger\hat{c}_{\ell'} \hat{\cal{R}})$, and 
 $\hat{\rho}_c$ has the elements  $\rho_{k,\ell}={\rm Tr}(\hat{b}_k^\dagger\hat{c}_{\ell} \hat{\cal{R}})$.  It can be  shown from Eq.~(\ref{Master_full}) that the introduced SPDMs obey the following system of  the three coupled  equations
\begin{align}
& \frac{\partial \hat{\rho}_{\rm s}}{\partial t}=-i[\widehat{H}_{\rm s}, \hat{\rho}_{\rm s}]
-i\epsilon(\widehat{V}^{\dagger}_1\hat{\rho}_{\rm c}-\hat{\rho}_{\rm c}^{\dagger}\widehat{V}_1), \label{one} \\
& \frac{\partial \hat{\rho}_{\rm c}}{\partial t}=-i\widehat{H}_{\rm r}\hat{\rho}_{\rm c}
+i\hat{\rho}_{\rm c}\widehat{H}_{\rm s}
-\frac{\gamma}{2}\hat{\rho}_{\rm c}
-i\epsilon(\widehat{V}_1\hat{\rho}_{\rm s}-\hat{\rho}_{\rm r}\widehat{V}_1), \label{two} \\
& \frac{\partial \hat{\rho}_{\rm r}}{\partial t}=-i[\widehat{H}_{\rm r}
,\hat{\rho}_{\rm r}]
-i\epsilon(\widehat{V}_1\hat{\rho}_{\rm c}^{\dagger} -\hat{\rho}_{\rm c}\widehat{V}^{\dagger}_1)
+\gamma(\hat{\rho}^{\scs (0)}_{ \rm r}-\hat{\rho}_{\rm r}) \;, \label{three}
\end{align}
where $\widehat{H}_{\rm{s}}$ is the single particle Hamiltonian of the chain,
\begin{equation}\label{Ham_single_sys}
\widehat{H}_{\rm{s}}=-\frac{J_{\rm s}}{2}\sum_{\ell=1}^{L-1}\left( |1\!+\!\ell\rangle\langle \ell|+\rm {h.c.}\right),
\end{equation}
$ \widehat{H}_{\rm{r}}$ the single particle Hamiltonian of the contact,
\begin{equation}\label{Ham_single_res}
\widehat{H}_{\rm{r}}=-J_{\rm{r}}\sum_{k=1}^M \cos\left(\frac{2\pi k}{M}\right)|k\rangle \langle k|,
\end{equation}
$\widehat{V}_{1}$ the coupling operator, 
\begin{equation}\label{V}
\widehat{V}_1=\frac{1}{2\sqrt{M}}\sum_{k=1}^M|k\rangle\langle \ell=1 | \;,
\end{equation}
and  $\hat{\rho}_r^{(0)}$ is the thermal SPDM of the carriers in the contact,
\begin{equation}
\hat{\rho}^{\scs (0)}_{ \rm r}=\sum_{k=1}^M\frac{| k \rangle\langle k |}{e^{-\beta[J_{\rm r}\cos(2\pi k/M))+\mu]}+1} \;.
\end{equation}
This set of equations can be easily  generalised onto the case of two contacts where the second contact is attached to last site of the chain.  When treating the latter case we shall change  notation for the contact SPDM from $\hat{\rho}_{\rm r}$ to $\hat{\rho}_\ell$, where $\ell$ takes the value $\ell=1$ for the left contact and $\ell=L$ for the right contact.

\subsection{The limit of small $\gamma$}
\label{sec3c}

Let us approximate Eqs.~(\ref{one}-\ref{three}) by the single equation of the form
\begin{equation}\label{Master_total}
\frac{\partial \hat{\rho}}{\partial t}=-i[\widehat{H}_{\rm},\hat{\rho}]
-\gamma\left(\hat{\rho}- \hat{\rho}_0 \right),
\end{equation}
where  $\hat{\rho}_0= \hat{\rho}^{\scs (0)}_{ \rm r}\oplus\hat{0}_{\rm s}$  ($\hat{0}_s$ is the zero matrix of the size $L\times L$) and $\widehat{H}$ is the total Hamiltonian, 
\begin{equation} \label{Ham_single}
\widehat{H}=
\left(
\begin{array}{cc}
\widehat{H}_{\rm{r}} & \epsilon\widehat{V}_1 \\
\epsilon\widehat{V}^{\dagger}_1 & \widehat{H}_{\rm{s}} 
\end{array}
\right) \;.
\end{equation}
This approximation is valid in the limit $\gamma\rightarrow0$ for the asymptotically large time where the matrix $\hat{\rho}$ is close to its stationary value. Notice that the limit $\gamma\rightarrow0$ implies that the system (\ref{Ham_single}) relaxes to its stationary value as the whole and, hence, there is no way to obtain the equation for $\hat{\rho}_{\rm s}$ in the closed-form.

\begin{figure}[t]
\includegraphics[width=8.5cm,clip]{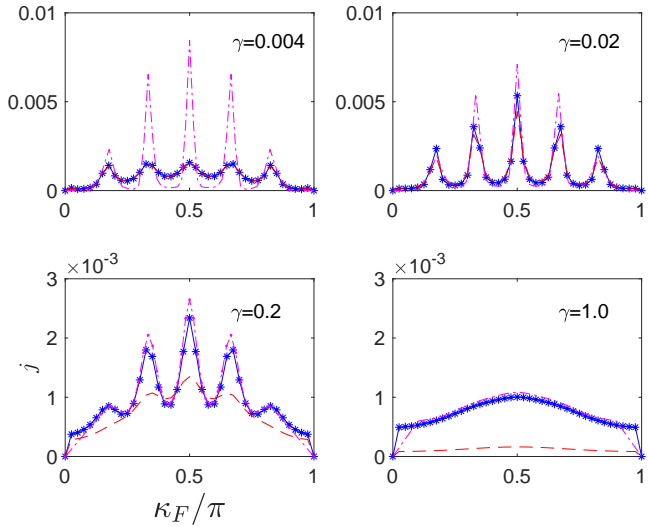}
\caption{Comparison of the stationary current calculated on the basis of Eq.~(\ref{d8}), dashed red lines, and on the basis of   Eq.~(\ref{a2}), dash-dotted magenta lines, with the result obtained on the basis of the original model, blue symbols.}
\label{fig3}
\end{figure}

The introduced Eq.~(\ref{Master_total}) can be solved analytically by using the eigenstates $|\Psi_n\rangle$,
\begin{equation}
\widehat{H}|\Psi_n\rangle={\cal E}_n|\Psi_n\rangle \;,
\end{equation}
of the total Hamiltonian (\ref{Ham_single}).  In particular, the stationary total density matrix is given by the equation \cite{77} 
\begin{equation}
\label{d7}
\hat{\rho}=\sum_{n,m}  \frac{\gamma\langle \Psi_n| \hat{\rho}_0 |\Psi_m\rangle}{\gamma+i({\cal E}_m-{\cal E}_n)}
 |\Psi_n\rangle\langle \Psi_m | \;.
\end{equation}
In the case of two contacts this equation determines the stationary current across the chain through the relation
%
\begin{equation}
\label{d8}
j=\sum_{p>0}\sum_{n} \frac{\gamma\langle \Psi_n| \hat{\rho}_0 |\Psi_{n+p}\rangle}{\gamma+i({\cal E}_{n+p}-{\cal E}_n)}
\langle \psi_{n+p} | \hat{j} |\psi_n \rangle  \;,
\end{equation}
where $| \psi_n\rangle$ is the part of the total wave function $| \Psi_n\rangle$ which resides in the chain and $\hat{j}$ the current operator,
\begin{equation}
\label{d9}
\hat{j}=-\frac{J_{\rm s}}{2i}\frac{1}{L-1}\sum_{\ell=1}^{L-1}\left( |\ell+1\rangle\langle \ell|-\rm {h.c.}\right) \;.
\end{equation}
In Fig.~\ref{fig3} we compare the stationary current calculated on the basis of the original master equation with that obtained on the basis of Eq.~(\ref{d8}). A good agreement for small $\gamma \le 0.02$ is noticed. 

\begin{figure}
\includegraphics[width=8.5cm,clip]{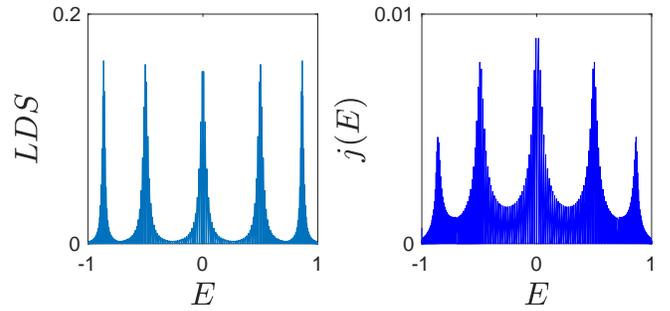}
\caption{The chain local density of states, the left panel, and the function (\ref{d10}), the right panel}
\label{fig4}
\end{figure}

Eq.~(\ref{d8}) also provides a simple explanation for the observed resonant structure of the current. In fact, this structure is already reproduced if we keep in Eq.~(\ref{d8}) only one term with $p=1$. (The term with $p=0$ vanishes due to the wave function symmetry.) Thus, the transmission peaks are due to the property of the current matrix elements which we shall characterise by the function
\begin{equation}
\label{d10}
j(E)=\sum_n \delta(E-{\cal E}_n) \langle \psi_n | \hat{j} |\psi_{n+1} \rangle .
\end{equation}
By definition, the function (\ref{d10}) is close to the local density of state,
\begin{equation}
\label{d11}
LDS=\sum_n \delta(E-{\cal E}_n) \langle \psi_n |\psi_n \rangle \;,
\end{equation}
which lies behind the concept of the level broadening  in the solid-state physics approach and which is directly related to the transmission probability $|t(E)|^2$. For the parameters of Fig.~\ref{fig3} the function $j(E)$ and the local density of states are plotted in Fig.~\ref{fig4}. In principle, by using the Green function one can extend the analysis of Eq.~(\ref{d8}) further \cite{Datt95}, however, this goes beyond our aim which is to identify the validity regions of the different approaches. It is seen in Fig.~\ref{fig3} (see also the dashed lines in Fig.~\ref{fig2}) that the approach based on the scattering theory underestimate the current for large $\gamma$.

\subsection{Non-Markovian master equation}
\label{sec3b}

In this subsection we discuss the quantum optics approach based on the non-Markovian master equation for the carriers in the chain.  We shortly come back to Eqs.~(\ref{one}-\ref{three}) which refer to the case of the single contact. From Eq.~(\ref{two}) we find the formal solution with the initial condition $ \hat{\rho}_{\rm c}(0)=0$
\begin{equation}\label{c}
\hat{\rho}_{\rm c}\!=\!i\epsilon\!\int\limits_{0}^{t}d\tau
e^{\frac{\gamma}{2}(\tau\!-\!t)}\widehat{U}^{\dagger}_{\rm r}(t\!-\!\tau)\!\left[
\hat{\rho}_{\rm r}(\tau)\widehat{V}_1\!-\!\widehat{V}_1\hat{\rho}_{\rm s}(\tau)\right]
\!\widehat{U}_{\rm s}(t\!-\!\tau),
\end{equation}
where $\widehat{U}_{\rm s,r}(t)=\widehat{\exp}(-i\widehat{H}_{\rm s,r}t)$ are the evolution operators. Employing the Born approximation, i.e., substituting $\hat{\rho}_{\rm r}(\tau)$ by $\hat{\rho}_{\rm r}^{(0)}$ in Eq.~(\ref{c}) and then substituting this equation into  Eq.~(\ref{one})  one obtains the non-Markovian master equation for the carriers in the chain,
\begin{equation}\label{master_reduced_one}
\frac{\partial \hat{\rho}_{\rm s}}{\partial t}=-i[\widehat{H}_{\rm s},\hat{\rho}_{\rm s}]
+\epsilon^2\left(\widehat{L}_1+
\widehat{L}_1^{\dagger}\right),
\end{equation}
where
\begin{equation}\label{Ldag}
\widehat{L}_1\!=\!\int\limits_{-t}^{0}d\tau e^{\frac{\gamma}{2}\!\tau}
\widehat{V}^{\dagger}_1\widehat{U}_{\rm r}^{\dagger}(\tau)\left[
\hat{\rho}_{\rm r}^{(0)}\widehat{V}_1\!-\!\widehat{V}_1\hat{\rho}_{\rm s}(\tau\!+\!t)\right]\widehat{U}_{\rm s}(\tau) \;,
\end{equation}
and where, to stress the memory  effect, we change the integration limits. Next, by taking the limit $M\rightarrow\infty$  we obtain 
 \begin{equation}\label{Ldag2}
\widehat{L}_1\!=\!\frac{| 1\rangle\langle 1|}{4} \!\int\limits_{-t}^{0}\!d\tau
e^{\frac{\gamma}{2}\tau}\!
\left[{\cal J}_{\rm F}(J_{\rm r}\tau)
\widehat{\mathbb I}_{\rm s}\!-\!{\cal J}_{0}(J_{\rm r}\tau)\hat{\rho}_{\rm s}(\tau\!+\!t)
\!\right]\widehat{U}_{\rm s}(\tau),
\end{equation}
where ${\cal J}_0$ is the zeroth order Bessel function of the first kind, $\widehat{\mathbb I}_{\rm s}$ is the identity matrix of the size $L\times L$, and
\begin{equation}\label{J_Fermi}
{\cal J}_{\rm F}(J_{\rm r} t)=\frac{1}{2\pi}\int\limits_{-\pi}^{\pi}d\kappa\frac{e^{-iJ_{\rm r}\cos(\kappa)t}}
{e^{-\beta[J_{\rm r}\cos(\kappa)+\mu]}+1} \;.
\end{equation}

In the case of two contacts the above procedure results in the equation
\begin{equation}\label{master_reduced}
\frac{\partial \hat{\rho}_{\rm s}}{\partial t}=-i[\widehat{H}_{\rm s},\hat{\rho}_{\rm s}]
+\epsilon^2\sum_{\ell=1,L}\left(\widehat{L}_{\ell}+ \widehat{L}_{\ell}^{\dagger}\right),
\end{equation}
where  the operator $\widehat{L}_L$ has the similar to Eq.~(\ref{Ldag2}) form with the projection operator $| 1\rangle\langle 1|$ substituted by $| L\rangle\langle L|$ and generally different value of the chemical potential in Eq.~(\ref{J_Fermi}). Equation (\ref{master_reduced})   together with Eq.~(\ref{Ldag2}) constitute the non-Markovian master equation for the fermionic transport \cite{remark2}. Notice the key role of $\gamma$ in Eq.~(\ref{Ldag2}) -- since the Bessel function at large $t$ decays as $1/\sqrt{t}$ the integral in Eq. (\ref{Ldag2}) is only convergent with non-zero $\gamma$.

To check the obtained non-Markovian  master equation we solve it numerically and compare the result  with that obtained on the basis of the original master equation where we do not take the limit $M\rightarrow\infty$ and do not a priori assume the validity of the Born approximation.   It is seen in Fig.~\ref{fig3}  that the non-Markovian master equation well reproduces the resonant structure for the stationary current  with nice quantitive  agreement in regions II and III.

The obtained master equation can be elaborated further if $\Delta \mu \ll \mu$ and one considers the low-temperature limit $\beta\rightarrow\infty$. The former condition justifies the ansatz
\begin{equation}
\hat{\rho}_{\rm s}=\hat{\rho}_{\rm s}^{\scs (0)}+\Delta\mu\hat{\rho}_{\rm s}^{\scs (1)}.
\end{equation}
where $\hat{\rho}_{\rm s}^{(0)}$ is the equilibrium density matrix for $\Delta \mu=0$ which does not support the directed current. The latter condition justifies the relation
\begin{equation}
\lim_{\beta\rightarrow\infty}n(E,\mu+\Delta\mu)=\theta(\mu-E)+{\Delta\mu}\delta(E-\mu),
\end{equation} 
where $\theta$ is the Heaviside function. Using these approximations one finds from Eq.~(\ref{master_reduced}) 
\begin{equation}\label{master_Land}
\frac{\partial\hat{\rho}_{\rm s}^{\scs (1)}}{\partial t}=-i[\widehat{H}_{\rm s},\hat{\rho}_{\rm s}^{\scs (1)}]
+\epsilon^2 \sum_{\ell=1,L}\left(\widehat{\Delta}_{\ell}+\widehat{\Delta}_{\ell}^{\dagger}\right).
\end{equation} 
where
\begin{align}\label{delta}
\widehat{\Delta}_{\ell}\!=\!\frac{|\ell\rangle\!\langle\ell|}{4}  \!\int\limits_{-t}^{0}\!d\tau
e^{\frac{\gamma\tau}{2}}\!\left[\!d(\mu)\delta_1^{\ell} e^{i\mu\tau}\widehat{\mathbb I}_{\rm s}
\!-\!{\cal J}_{0}(J_{\rm r}\tau)\hat{\rho}_{\rm s}^{\scs (1)}(\!\tau\!+\!t)\!\right]\!\widehat{U}_{\rm s}(\!\tau\!),
\end{align} 
with $d(\mu)$ being the contact density of states
\begin{equation}\label{DOS}
d(\mu)=
\left\{
\begin{array}{cl}
\frac{J_{\rm r}}{\pi\sqrt{J_{\rm r}^2-\mu^2}} & \ {\rm if} \ |J_{\rm r}|>|\mu| \\
0  & \ {\rm if} \ |J_{\rm r}|<|\mu| 
\end{array}
\right. \;.
\end{equation}
Finally, from Eq.~(\ref{master_Land}) we obtain the algebraic equation for the stationary $\hat{\rho}_{\rm s}^{\scs (1)}$.  In the chain eigenbasis $|\Phi_n\rangle$, 
\begin{equation}
\label{a1}
\widehat{H}_{\rm s}|\Phi_n\rangle=E_n |\Phi_n\rangle \;,
\end{equation}
it reads
\begin{equation}
\label{a2}
i\left[ H_{\rm s},\hat{\rho}_{\rm s}^{\scs (1)}\right]
+\frac{\epsilon^2}{4} \left[ (C_1 + C_L)\hat{\rho}_{\rm s}^{\scs (1)} B + h.c \right]
=\frac{\epsilon^2}{4}\left(C_1 A + h.c.\right)  \;,
\end{equation}
where $H_{\rm s}$ is the diagonal matrix of the eigenvalues $E_n$, $B$ the diagonal matrix with the elements
\begin{equation}
\label{a3}
B_{n,n}=\frac{1}{\sqrt{J_{\rm r}^2-(E_n+i\gamma/2)^2}}  \;,
\end{equation}
$A$ the diagonal matrix with the elements
\begin{equation}
\label{a4}
A_{n,n}=\frac{ d(\mu) }{\gamma/2+i(\mu-E_n)}  \;,
\end{equation}
and $C_1$ and $C_L$ are determined by the eigenmodes  of the isolated chain,
\begin{equation}
\label{a5}
C_{n,m}= \langle \Phi_n | \ell \rangle \langle \ell | \Phi_m\rangle \;,\quad \ell=1,L  \;.
\end{equation}
It follows from Eq.~(\ref{a2}) that the matrix  $\hat{\rho}_{\rm s}^{\scs (1)}$  (which is proportional to r.h.s. of the equation) crucially depends on the value of the chemical potential due to the resonance-like dependence of the matrix $A$ on the chemical potential $\mu$.  

The advantage of the discussed algebraic approach (as well as the algebraic approach in Sec.~\ref{sec3c})  is that it allows us to predict the stationary current without simulating  the system dynamics, which reduces the computational time by orders of magnitude. For large $\gamma$ the predictions based on Eq.~(\ref{a2}) are plotted in Fig.~\ref{fig2} by the dash-dotted lines. Taken into account that Eq.~(\ref{a2}) involves even more approximations than the non-Markovian master equation (\ref{master_reduced}), the agreement with results of the straightforward numerical simulation of the system dynamics is quite satisfactory.

\subsection{Markovian master equation}
\label{sec3a}

The Markov approximation assumes that one can neglect the memory effects. It is justified if $\hat{\rho}_s(t)$ is a slowly varying matrix in the  time scale $1/\gamma$. Then, by using the general relation for the slowly varying function,
\begin{equation}
\int\limits_{0}^{t}d\tau
e^{\frac{\gamma}{2}\tau } {\cal A}(\tau+t) = \frac{2}{\gamma}{\cal A}(t) \;,
\end{equation}
Eq.~(\ref{Ldag2}) simplifies to
\begin{equation}\label{Markov1}
\widehat{L}_\ell\!=\!\frac{| \ell\rangle\langle \ell|}{2\gamma} 
\left[\bar{n}_\ell \widehat{\mathbb I}_{\rm s}\!-\!\hat{\rho}_{\rm s}(t)\!\right] ,
\end{equation}
where we take into account that ${\cal J}(0)=1$ while ${\cal J}_{\rm F}(0)$ equals to the mean occupation number $\bar{n}_\ell$ of the ring sites according to Eq.~(\ref{J_Fermi}).  Substituting Eq.~(\ref{Markov1}) into Eq.~(\ref{master_reduced}) we obtain
\begin{equation}\label{BM}
\frac{\partial \hat{\rho}_{\rm s}}{\partial t}\!=\!-i[\widehat{H}_{\rm s},
\hat{\rho}_{\rm s}]\!-\tilde{\gamma}\sum_{\ell=1,L}\!
\left(\frac{1}{2}\left\{|\ell \rangle\langle \ell|,\hat{\rho}_{\rm s} \right\}
\!-\!\bar{n}_{\ell}|\ell \rangle\langle \ell|\right)\!,
\end{equation}
where
\begin{equation}
\tilde{\gamma}=\frac{\epsilon^2}{\gamma} \;.
\end{equation}
It follows from Eq.~(\ref{BM}) that the characteristic time scale for $\hat{\rho}_{\rm s}(t)$ is determined by the inverse effective relaxation constant $1/\tilde{\gamma}$. Thus, the Markov approximation requires $\gamma\gg\epsilon$.

The obtained Markovian master equation Eq.~(\ref{BM}) coincides with the equation for SPDM of the open Fermi-Hubbard model, which in the case of spinless Fermions reads
\begin{equation}\label{BH1}
\frac{\partial \widehat{{\cal R}}_s}{\partial t}=-i[\widehat{{\cal H}_s},  \widehat{{\cal R}}_s]+
\tilde{\gamma}\sum_{\ell=1,L}\left(\widehat{{\cal L}}^{\scs (g)}_{\ell}+\widehat{{\cal L}}^{\scs (d)}_{\ell}\right) \;,
\end{equation}
where $\widehat{{\cal H}}_s$ is given in Eq.~(\ref{Ham_sys}) and the drain and gain Lindblad operators acting on the edge sites of the chain have the form
\begin{equation}
\widehat{{\cal L}}^{\scs(d)}_{\ell}=\frac{\bar{n}_{\ell}-1}{2}
\left(\hat{c}_{\ell}\hat{c}_{\ell}^{\dagger}\widehat{\cal R }-2\hat{c}_{\ell}^{\dagger}\widehat{\cal R }\hat{c}_{\ell}
+\widehat{\cal R }\hat{c}_{\ell}\hat{c}_{\ell}^{\dagger} \right) \;,
\end{equation}
and 
\begin{equation}
\widehat{{\cal L}}^{\scs{(g)}}_{\ell}=-\frac{\bar{n}_{\ell}}{2}
\left(\hat{c}_{\ell}\hat{c}_{\ell}^{\dagger}\widehat{\cal R }-2\hat{c}_{\ell}^{\dagger}\widehat{\cal R }\hat{c}_{\ell}
+\widehat{\cal R }\hat{c}_{\ell}\hat{c}_{\ell}^{\dagger} \right) \;.
\end{equation}
The open Fermi-Hubbard model was analysed earlier in Ref.~\cite{114} with the following result for the mean current
\begin{equation}\label{BH2}
j = \frac{J_{\rm s}^2\tilde{\gamma}}{J_{\rm s}^2 + \tilde{\gamma}^2}  \frac{\bar{n}_{1}-\bar{n}_{L}}{2} \;.
\end{equation}
Since Eq.~(\ref{BH2}) involves only the mean density of carriers in the contacts it approximates the exact results only in the limit $\gamma\rightarrow\infty$ where the stationary current does not show any resonant structure.

\section{Coherence of the transporting states}
\label{sec4}

\begin{figure}
\includegraphics[width=8.5cm,clip]{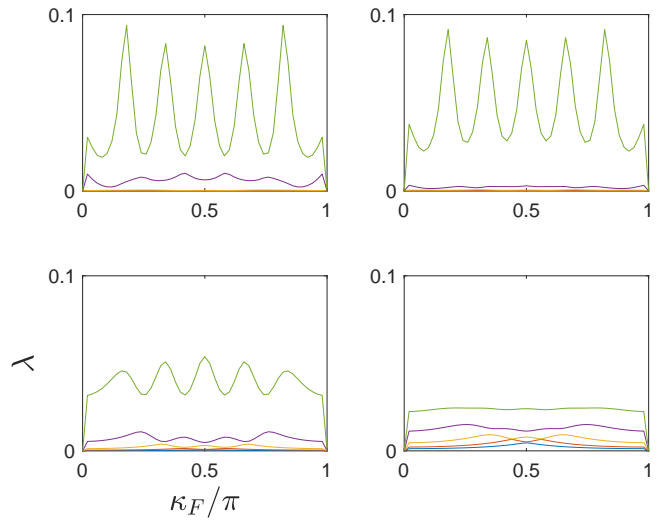}
\caption{Eigenvalues of stationary matrix $\hat{\rho}_{\rm s}^{\scs (1)}$ as the function of $\kappa_F$. The system parameters are the same as in Fig.~\ref{fig3}.}
\label{fig5}
\end{figure}

In the end of Sec.~\ref{sec3c} we introduce the density  matrix  $\hat{\rho}_{\rm s}^{\scs (1)}$ which characterises the stationary current across the chain in the liner response regime (i.e., determines the conductance). In what follows we shall term this matrix as the transporting state. It is interesting to address the question of how close is the transporting state to a pure state. In Fig.~\ref{fig5} we plotted eigenvalues of $\hat{\rho}_{\rm s}^{\scs (1)}$ for four different values of the relaxation constant $\gamma$ which we used earlier in Fig.~\ref{fig3}.  By comparing Fig.~\ref{fig5} and Fig.~\ref{fig3} we conclude that the transporting state is close to a pure state  (which has to have the single non-zero eigenvalue) only in the interval of $\gamma$ where the stationary current shows the resonant structure. Moreover, even for these $\gamma$  coherence of the transporting state depends on the value of the chemical potential $\mu$. Namely, the state is more coherent for $\mu$ corresponding to the transmission peaks and essentially less coherent for $\mu$ corresponding to the transmission deeps.   
This result stresses the main difference of the discussed master equation approach with the Landauer-like  approaches which implicitly assumes that the transport state of the fermionic carriers in the chain is the pure state given by the Bloch wave with the Fermi quasimomentum.

\section{Conclusion}
\label{sec5}

We revisite the problem of  the quantum transport of fermionic particles across the tight-binding chain connected to two contacts.  The analysis is performed by using the simple model introduced in our earlier work \cite{120} where the  contacts are modelled by the tight-binding rings of  arbitrary size. The coupling between the chain and contacts is controlled by the parameter $\epsilon$ and the contacts are characterized by their chemical potentials $\mu$ and the temperature $1/\beta$. The mathematical framework of the model is the master equation for the total density matrix of the composed system chain plus contacts which involves the important physical parameter $\gamma$ -- the rate at which the isolated contacts relax to the thermodynamic equilibrium. Although in reality the rate  $\gamma$ correlates with  the contact temperature (indeed, it looks unfeasible to have low rates at high temperatures and visa versa), in the work we consider $\gamma$ and $\beta$ as independent parameters.

We calculate the current in the chain by using the three different approaches -- the Markovian and non-Markovian master equations for the reduced density matrix of the carriers in the chain, and by the Landauer-like approach for the quantum transport.  We discuss in detail each of the listed approaches and identify the regions of their validity. In particular, it is found that the non-Markovian master equation approach (which is a representative of quantum optics approaches) and the Landauer approach (which is representative of the solid-state physics approaches) do reproduce the resonant structure of the stationary current. However, the former approach overestimates it in the region of small $\gamma$, while the latter approach underestimates the current in the region of large $\gamma$. For moderate $\gamma$, where oscillations of the stationary current $j=j(\mu)$ have maximal amplitude, both the solid-state physics and quantum optics approaches give essentially the same result.


We acknowledge financial support from Russian Science Foundation through Grant No. 19-12-00167.



\begin{thebibliography}{99}
\bibitem{remark1}
We mention that this statement does not imply the absence of entanglement between the system (the chain) and reservoirs (the contacts). On the contrary, the entanglement is always there and it is responsible for the decoherence effect of reservoirs \cite{117}.
\bibitem{remark2}
Eq.~(\ref{master_reduced}) also describes the transport of non-interacting Bose particles if one uses the Bose-Einstein distribution  instead of the Fermi-Dirac distribution Eq.~(\ref{Fermi}). 
\end{thebibliography}
\end{document}